\documentclass[aps,pre,preprint,groupedaddress,twocolumn,nofootinbib]{revtex4}
\usepackage{graphicx}

\begin{document}


\title{Improving the Global Fitting Method on Non-Linear Time Series Analysis}


\author{L.M.C.R. Barbosa}
\affiliation{Universidade Federal Fluminense, Instituto de F\'isica
Campus da Praia Vermelha, Gragoat\'a, Niter\'oi, RJ
CEP: 24210-310}

\author{L.G.S. Duarte, C.A. Linhares and
L.A.C.P. da Mota}
\email[]{lduarte@dft.if.uerj.br, linhares@dft.if.uerj.br, damota@dft.if.uerj.br}
\affiliation{Universidade do Estado do Rio de Janeiro, Instituto de F\'{\i}sica, Depto. de F\'{\i}sica Te\'orica, 20559-900 Rio de Janeiro -- RJ, Brazil}


\date{\today}

\begin{abstract}
 In this paper, we are concerned with improving the forecast
capabilities of the Global approach to Time Series. We assume that the
normal techniques of Global mapping are applied, the noise reduction is
performed, etc. Then, using the mathematical foundations behind such
approaches, we propose a method that, without a great computational
cost, greatly increase the accuracy of the corresponding forecasting.
\end{abstract}


\maketitle

\section{Introduction \label{intro}}

For any observed system, physical or otherwise, one generally
wishes to make predictions on its future evolution. Sometimes,
very little is known about the system. Possibly, the dynamics
behind the phenomenon being studied is unknown, and one is given
just a time series of one (or a few) of its parameters. Therefore,
performing a time-series analysis is the best one can do in order
to learn the properties of the phenomenon. Its relevance may be
gauged by the existence of extensive studies in a great diversity
of branches of knowledge, in physics as well as in economics and
the stock exchange, meteorology, oceanography, medicine, etc.

A time series is normally taken as a set of numbers that are the possible
outcome of measurements of a given quantity, taken at regular intervals. In
reality, however, the assumption that the time series reflects in some way
the underlying dynamics of the systems is worsened by the fact that the
measured data usually contain irregularities. These may be due to a random
external influence on a linear system, a noise (induced possibly by the
measuring apparatus or other sources of contamination) which gets mixed with
the desired information, thereby hiding it. But it may well be that they
appear as a manifestation of low-dimensional deterministic chaos resulting
from an intrinsic nonlinear dynamics governing the quantity under study
(over which a random noise may also be superimposed), with the
characteristic sensitivity to initial conditions.

If the time series is the only source of information on the
system, prediction of the future values of the series requires a
modelling of the system's (perhaps nonlinear) dynamical law
through a set of differential equations or through discrete maps.
However, it is even possible that we do not know whether the
measured quantity is the only relevant degree of freedom
(frequently it is not) of the dynamical problem, nor how many of
them there are.

Both noise-contaminated linear and nonlinear systems have
nevertheless been studied with success employing statistical
tools, chaos-theory concepts, together with time-series analysis
\cite{AbarbanelN,caos}. Given a time series, one should ask first
whether it represents a causal process or whether it is
stochastic. Tools have been developed to decide upon this
fundamental question (the most common ones are spectral analysis,
Lyapunov characteristic exponents and correlation functions, see
\cite{Kantz,Abarbanel}). In the case of a series originated from a
low-dimensionality chaotic dynamics, traditional linear methods of
analysis are not adequate, but an analysis apparatus was devised
for applications to such nonlinear systems \cite{Kantz,Abarbanel}
and we will not be concerned with stochastic processes in this
paper.

Methods for dealing with nonlinear time series fall mainly into two
categories: local or global methods. Local methods are based on the
assumption that, while in the long run nearby trajectories on the phase
space diverge considerably, they stay within the same neighborhood for a
while. One may conjecture that to predict the next step in a time series, a
good indication should come from the previous visits the system had made to
the phase space neighborhood containing the ``last point'' of the series. An
average of the behavior of the system for neighboring points, with a
minimization of the distance in the phase space between them, gives good
results for the next-step forecasting.

Global methods, on the other hand, postulate a functional form for the
dynamics to be valid for any time. Usually one considers polynomials of a
suitable degree and one should devise a convenient way to estimate its
coefficients.
In this paper, we are going to concentrate in the global approach and,
actually, we will start from the global mapping itself, i.e., we are not
going to be concerned with how the global mapping was generated (there
are many standard approaches to do it) and we will not deal with noise
reduction either (such considerations are important when determining the
mappings, etc.). We will focus on a new method to, from any standard
mapping one might have, improve the forecasting using it, without having
to pay a very high computational price.

Nonlinear analysis of Time Series relies not on the original maps of the
dynamic system, but on its {\em time-delay reconstruction}. All discussions on
the nonlinear treatment of Time Series make use of this reconstruction
scheme. There are already classical references dealing with the
subject \cite{AbarbanelN,Kantz,Abarbanel,Abarbanel2,Hegger,Schreiber,Kugiumtzis}. This method allows one to reconstruct the phase space of the
system with reasonable accuracy, using the information contained in the
series only.

Lorenz \cite{Lorenz} showed that dynamic systems of low dimensionality
could present strange attractors on their phase spaces. Takens
\cite{Takens} proposed a method to reconstruct such phase spaces from
the knowledge of a Time Series obtained from the system. He demonstrated that the
original attractor and the reconstructed one are characterized by the
same asymptotic properties and topological characteristics \cite{topo}.
So, if we want to analyze the properties of the corresponding attractor
of the system we have to reconstruct it.

In \cite{Takens}, Takens used a method to reconstruct the phase space.
Vectors $\overrightarrow{\xi_i}$ (with dimension ``m'') are
reconstructed from the Time Series ${x_i}$ where
$x_i=x(t_i)$,$i=1,...,N$ as follows:
\begin{equation}
\overrightarrow{\xi_i}=\{x(t_i),x(t_i+p),...,x(t_i+(m-1)p)\}
\end{equation}
where $m$ is the embedding dimension and $p$ is the time lag (for
definitions, see  \cite{embd}).
Based on the trajectories of the reconstructed attractor, we can
study various topological invariants of the system such as the
Lyapunov exponents, the generalized entropies \cite{topo}, etc. We
can also extract the underlying dynamics via a global modelling of
the system. For example, one can try to obtain a low order Taylor
series expansion for the system, thus obtaining a global mapping
representing the system. We can use this mapping to perform a
forecast of entries we ignore, i.e., in the future\footnote{One
can also do forecasting in a local version via analyzing the
behavior of close vectors (to the one just before the one to be
predicted) in order to estimate the next (unknown) entry (see
\cite{AbarbanelN}).}.


\section{An Algorithm to Improve the Global Forecasting}
\label{algo}

\subsection{Stating the problem}

Suppose that the system can be modelled by a set of differential
equations of low dimensionality. What we would like to obtain is
some kind of global map that, given any point of the phase space,
could calculate a subsequent point of the trajectory. If we have
known the set of differential equations (SDE) that models the
system, we could find a solution (starting from an initial
condition) by making a numerical integration through some map
obtained from the SED (probably a Runge-Kutta map, a Taylor series
one or an expansion in some function basis). For practical
purposes (computers can not work with the infinity) a truncation
must occur at some order of the series expansion. However, if the
truncation order is low, we can run away from the real solution in
a few time steps (even if each time step is very small). For
chaotic systems it is not used (in general) a Runge-Kutta
expansion of degree less than four. This implies that the map
generated present polynomials of high degree. Let's exemplify
using one of the simplest chaotic system that exists, the Lorenz
system:

\begin{eqnarray}
\label{sistLorenz}
\dot{x}_1 & = & \sigma\,(x_2-x_1), \nonumber \\
\dot{x}_2 & = & - x_2 - x_1\,x_3 + R\,x_1,  \; \\
\dot{x}_3 & = & x_1\,x_2 - b\,x_3, \nonumber
\end{eqnarray}
\noindent
where $\sigma$, $R$ and $b$ are parameters and the system presents
chaotic behavior for $R > 24,74$.

Why {\em one of the simplest}? Notice that this system possesses the
minimum number of autonomous\footnote{The time does not appear
explicitly.} differential equations permeating chaos: three\footnote{In
two dimensions we can not have chaos because the trajectories can not
cross.}. Besides that, chaos is a phenomenon that only takes place in
non-linear systems, and the smallest piece of non-linearity that we can add
to a linear system in order to turn it non-linear is a quadratic term.

Observe that the Lorenz system presents only two non-linear quadratic
terms. Even in this simple case, as we will show, a Taylor series expansion of fourth
order, will lead to a map of fifth degree in three variables.

Consider the following initial condition:
$x_1(0)={x_1}_0,\,x_2(0)={x_2}_0,\,x_3(0)={x_3}_0$. We can expand the
corresponding solution as:
\begin{equation}
\label{solLorenz}
x_i = \phi_i(t) = \phi_i(0) + \frac{d\phi_i}{dt}(0)\,t +
\frac{d^2\phi_i}{dt^2}(0)\,\frac{t^2}{2!} + \cdots.
\end{equation}
\noindent Since the system is defined by the equations
$\frac{dx_i}{dt} = f_i(\vec{x}),$
\noindent we have that
$\dot{x_i}=\dot{\phi_i}=f_i$
\footnote{ Where $\dot{u}$ represents
$\displaystyle{\frac{du}{dt}}$.}, implying that:
\begin{equation}
\label{map1Lorenz}
\frac{d}{dt}\left(\frac{d\phi_i}{dt}\right) = \frac{df_i}{dt} =
\sum^3_{j=1} \, \frac{\partial f_i}{\partial x_j}\frac{dx_j}{dt} =
\sum^3_{j=1} \, \frac{\partial f_i}{\partial x_j}f_j.
\end{equation}

We can notice that, for the case of the Lorenz system, this process will
increase the degree of the polynomials forming the mapping by one for
each order\footnote{Since the highest degree present in the functions
$f, g$ and $h$ is quadratic, the derivatives (present on
(\ref{map1Lorenz})) are, at maximum, first degree polynomials.}.
So, the mapping corresponding to the forth order Taylor expansion is, at
maximum, formed by fifth degree terms. A polynomial mapping of fifth
degree implies a total of $168$ coefficients. Please remember that, as
mentioned, this is for one of the simplest chaotic dynamic system cases
(i.e., three-dimensional and only two non-linear (quadratic) terms).

It is important to notice that, in Time Series analysis, we do not have
the dynamic system to begin with. We, of course, will consider that
there is such a system behind the series and we will look for
determining it. With the explanations above, we hope to have made it
clear that, even if the underlying system is as simple as the Lorenz's
one, we will already have to face a great computational task (if one
wants to use forth order expansions - generally the minimum accuracy
necessary for practical purposes) of determining the $168$ coefficients.
With more detail, using the Lorenz system as a model for the Global
Fitting scheme, let us suppose that we have a Time Series produced from
this system (for instance, take one of the coordinates of the system).
After the usual phase space reconstruction \cite{Abarbanel}, say we want
to have a fourth order mapping (for the reconstructed system) with the
same accuracy that could be found on the fourth order Taylor expansion
for the Lorenz system. We would have to employ some minimization
technique to determine $168$ coefficients. In practice, this is a very
high number making the whole procedure computationally expensive.

So, we are left with the hard choice of: either pay the computational
price mentioned above and be very patient or try and decrease the degree
of the mapping. Of course, there is no such thing as a free meal. The
price for the latter choice would be that the accuracy would decrease
(the corresponding Taylor expansion would be of lower order).

Therefore, despite the fact that the global approach has many attractive
features, such as the fact that, once it is determined it is applicable
to the whole series\footnote{In the case of Local mappings, we have to
determine a mapping for each entry of the series.}, one sees that the
effective use of it can be difficult to achieve in practice. So, there
is a clear demand for procedures that can, without increasing the degree
of the global mapping, enhance the accuracy of such mappings. In the
next subsection, before introducing one such attempt, we will talk about mappings.

\subsection{Regarding Mappings}
\label{lie}

In order to clarify the central idea of our proposed algorithm, let us
make some comments and present some results concerning mappings
representing the solutions for SDEs.

\hspace \parindent
Consider the transformation group in $n$ variables:
\begin{equation}
\label{grupodetransf}
{x_i}^{\ast} = F_i(\vec{x},t),
\end{equation}
\noindent
where $t$ is the group parameter. From Lie's theory \cite{step,bluman,olver}, we know that
this group is the solution to a SDE defined by:
\begin{equation}
\label{eqdifdogrupo}
\dot{x_i} = f_i(\vec{x}),
\end{equation}
\noindent
where $f_i(\vec{x}) \equiv {\frac {\partial F_i}{\partial t}}\mid_{t=0}$
and $\dot{x_i} \equiv \frac {d x_i}{d t}$. Therefore, the transformation
group (\ref{grupodetransf}) (i.e., the solution to the dynamic system
(\ref{eqdifdogrupo})) can be obtained from the group generator defined
as the operator $X \equiv {\sum^n}_{i=1} \, f_i
\,\frac{\partial }{\partial x_i}$, as follows:
\begin{widetext}
\begin{equation}
\label{grupoSeriTaylor}
{x_i}^{\ast} = F_i(\vec{x},t) = x_i + t{\it
X}[x_i]+{\frac{t^{2}}{2!}} {\it X}^{2}[x_i]+ \cdots =
\sum_{k=0}^{\infty}{\frac{t^{k}}{k!}}{\it X}^{k}[x_i].
\end{equation}
\end{widetext}
\noindent
In this way, starting from a generic point $P_0$, with corresponding
coordinates $\vec{x}_{(P_0)}$, by choosing a time interval $\delta t$,
the transformation group (\ref{grupodetransf}) generates a mapping $M$
that takes a point on some given solution to the system and takes it to
another such point that corresponds to a group parameter increased by
$\delta t$
\begin{equation}
\label{mapaSolu}
{x_i}_{(P+1)} = F_i(\vec{x}_{(P)},\delta t) =
\sum_{k=0}^{\infty}{\frac{{\delta t}^{k}}{k!}}{\it X}^{k}[{x_i}_{(P)}].
\end{equation}
\noindent
In practice, the process of numerically solving the SDE can be summarized
by choosing a small time interval ($\delta t \ll 1$) and truncating the
series (\ref{mapaSolu}) at some order $\scriptstyle{N}$, thus obtaining
a mapping $\overline{M}$ given by:
\begin{equation}
\label{mapaSoluTrunc}
{\overline{x}_i}_{(P+1)} = \overline{F}_i(\vec{x}_{(P)},\delta t) =
\sum_{k=0}^{N}{\frac{{\delta t}^{k}}{k!}}{\it X}^{k}[{x_i}_{(P)}],
\end{equation}
\noindent
where ${\overline{x}_i}_{(P+1)}$ approaches ${x_i}_{(P+1)}$ when $\delta
t \rightarrow 0$. Defining the functions
$\delta^k \varepsilon_i$ as
\begin{widetext}
\begin{eqnarray}
\label{varep}
\left( \varepsilon_i(\vec{x}_{(P)})\right. & = &\left. \delta^0 \varepsilon_i(\vec{x}_{(P)})\right) \equiv  {x_i}_{(P+1)} - {\overline{x}_i}_{(P+1)} =
\sum_{k=N+1}^{\infty}{\frac{t^{k}}{k!}}{\it X}^{k}[{x_i}_{(P)}] \nonumber \\
\left( \delta \varepsilon_i(\vec{x}_{(P)})\right.   & = & \left. \delta^1 \varepsilon_i(\vec{x}_{(P)}) \right)  \equiv \varepsilon_i(\vec{x}_{(P+1)}) -
\varepsilon_i(\vec{x}_{(P)}) \nonumber \\
\delta^k \varepsilon_i(\vec{x}_{(P)}) & \equiv & \delta^{k-1} \varepsilon_i(\vec{x}_{(P+1)}) - \delta^{k-1} \varepsilon_i(\vec{x}_{(P)}),
\end{eqnarray}
\end{widetext}
\noindent
where (k=2,\ldots), one can notice that
\begin{equation}
\label{deltaep}
\delta \varepsilon_i(\vec{x}_{(P)}) = \sum_{j=1}^{n}
\frac{\partial \varepsilon_i(\vec{x}_{(P)})}{\partial x_j} \, \delta x_i +
\, O({\delta x_i}^{2})
\end{equation}
\noindent
and, generally,
\begin{eqnarray}
\label{deltakep}
\delta^{k+1}\!\varepsilon_i(\vec{x}_{(P)}) = \delta \, \delta^k\!\varepsilon_i(\vec{x}_{(P)}) = \nonumber \\
\sum_{j=1}^{n}
\frac{\partial \, \delta^k\!\varepsilon_i(\vec{x}_{(P)})}{\partial x_j} \, \delta x_i +
\, O({\delta x_i}^{2}).
\end{eqnarray}
\noindent
Since $\delta t \rightarrow 0$ implies that $\delta x_i \rightarrow 0$,
we can, using (\ref{deltakep}), enunciate the following result:
\begin{equation}
\label{Resultado}
\lim_{\delta t \rightarrow 0} \frac{\delta^{k+1} \varepsilon}{\delta^{k}
\varepsilon} = 0,
\end{equation}
\noindent
where $k$ is a positive integer.

In the next subsection, based on this important result, we will present
an algorithm that enhances the predictive power of global mappings for Time
Series.

\subsection{Mathematical Basis for the Algorithm}
\label{dalgo}

\hspace
\parindent
Based on the above result (\ref{Resultado}), we have produced an
algorithm that allows for improving the forecasting for the global
fitting of a Time Series.

As mentioned, we will suppose that the given Time Series is originated
from  phenomena that can be described by a low dimension dynamic
system ($S_0$). After the phase space reconstruction \cite{Takens}, we have a set of
vectors defining a set of points along a single trajectory  of the
reconstructed systems ($S_r$)\footnote{Takens \cite{Takens} has demonstrated that the
system $S_0$ and $S_r$ are topologically equivalent.}. As usual, what we would like to determine
is a global mapping $M$ that would (with infinite precision) represent the
solutions of the system $S_r$. But, of course, in practice, what we can
do is to produce a global mapping $\overline{M}$ through a procedure
involving a minimization process\footnote{In layman terms, what is done
is to adjust the coefficients of the polynomial mapping (of a certain
degree) to better reproduce the phase space points.}.
If the Mapping  $\overline{M}$ produces good forecasting
for the series,
that means that the coefficients present on $\overline{M}$ are close to
the analogous ones present on the mapping $M$ which can be represented by the infinite series
(\ref{mapaSolu}) (and, ideally, it would describe $S_r$ with absolute
precision). In that situation, we
would be in a similar position to the one presented on the last
subsection (where we had just a
truncated series because we knew the underlying SDE and could determine the Taylor expansion).
Why similar? In the ``real'' case we are dealing with now, we only have the series and have to
determine the mapping through a finite process and, therefore, the
coefficients would not be exactly the same as in the truncated expansion
of $S_r$. So, defining functions $\varepsilon_i$ and $\delta^k \varepsilon_i$
analogously to how we did in the last subsection, we would expect that (\ref{Resultado})
would be valid. Actually, in the real world, the inequality
\begin{equation}
\label{ResultadoReal}
{\delta^{k+1} \varepsilon} \ll {\delta^{k} \varepsilon}
\end{equation}
\noindent
is not valid for any positive integer $k$. The point is that, in actual calculations,
$\delta t$ would be a finite value (not infinitesimal) $\Delta t$. So, at some integer value
$\scriptstyle{K}$, the inequality (\ref{ResultadoReal}) would become
\begin{equation}
\label{ResultadoReal2}
{\Delta^{K+1} \varepsilon} \approx {\Delta^{K} \varepsilon}.
\end{equation}
\noindent
The above reasoning allows us to build an easily applicable algorithm:
Consider that we want to forecast the coordinate $x_i$ (where $i$ can
take any value from 1 to the dimensionality of the reconstructed system)
of a point $P+1$ that immediately follows a given point $P$. In order to
produce the mapping $\overline{M}$, we use a certain number $a+1$ of
points that precede the point $P+1$ (the points $P,\,P-1,\,P-
2,\,\ldots,\,P-a$). Using this mapping, we can forecast the $x_i$
coordinates for these $a+1$ points. Let us call these $a+1$ values
$\overline{x}_i$. From these, we can define the functions $\Delta^k
\varepsilon_i$ (analogously to the functions (\ref{varep}) in subsection
\ref{lie}).
\begin{eqnarray}
\label{varepReal}
\Delta^0 \varepsilon_i(\vec{x}_{(J)}) & \equiv & {x_i}_{(J)} - {\overline{x}_i}_{(J)} \nonumber \\
\Delta^1 \varepsilon_i(\vec{x}_{(J)}) & \equiv & \Delta^0 \varepsilon_i(\vec{x}_{(J)}) -
\Delta^0 \varepsilon_i(\vec{x}_{(J-1)}) \nonumber \\
\vdots & & \vdots \nonumber \\
\Delta^k \varepsilon_i(\vec{x}_{(J)}) & \equiv & \Delta^{k-1} \varepsilon_i(\vec{x}_{(J)}) -
\Delta^{k-1} \varepsilon_i(\vec{x}_{(J-1)}),  \nonumber \\
\vdots & & \vdots
\end{eqnarray}
\noindent
where $(k=0,\ldots)$ and $(J=P-a+k,\ldots,P)$. Using these definitions,
we can determine the values for $\scriptstyle{k}$ where we have
${\Delta^{\scriptstyle{k+1}} \varepsilon} \approx
{\Delta^{\scriptstyle{k}} \varepsilon}$\footnote{There is a finite range for the values for $k$
in which that happens. After a certain value, the $\Delta^{k-1}
\varepsilon_i$ start to diverge.} and, using this knowledge, we will see
that we can improve the forecasting generated by the mapping $\overline{M}$.
Let us clarify what we mean: if we want to forecast the value for the
coordinate $x_i$ of the point $P+1$, we may use the global mapping
$\overline{M}$ that would produce the forecast ${\overline{x}_i}_{(P+1)}$.
We know that ${x_i}_{(P+1)} - {\overline{x}_i}_{(P+1)} = \Delta^0
\varepsilon_i(\vec{x}_{(P+1)})$ and, therefore,
\begin{equation}
\label{pre0}
{x_i}_{(P+1)} = {\overline{x}_i}_{(P+1)} + \Delta^0
\varepsilon_i(\vec{x}_{(P+1)}).
\end{equation}
\noindent
Notice that we do not know the value for $\Delta^0
\varepsilon_i(\vec{x}_{(P+1)})$. But we know that $\Delta^1 \varepsilon_i(\vec{x}_{(P+1)}) =
\Delta^0 \varepsilon_i(\vec{x}_{(P+1)}) - \Delta^0
\varepsilon_i(\vec{x}_{(P)})$, implying that
\begin{equation}
\label{del1}
\Delta^0 \varepsilon_i(\vec{x}_{(P+1)}) = \Delta^0
\varepsilon_i(\vec{x}_{(P)}) + \Delta^1
\varepsilon_i(\vec{x}_{(P+1)}).
\end{equation}
\noindent
Let us examine this: we know the value for $\Delta^0
\varepsilon_i(\vec{x}_{(P)})$ (i.e., ${x_i}_{(P)} -
{\overline{x}_i}_{(P)}$) but we do not know $\Delta^1
\varepsilon_i(\vec{x}_{(P+1)})$. However, if $(P)$ and $(P+1)$ are
sufficiently close (such that ${\Delta^1 \varepsilon_i} \ll {\Delta^0
\varepsilon_i}$), we can expect that we will gain information when
substituting (\ref{del1}) into (\ref{pre0}) obtaining
\begin{equation}
\label{pre1}
{x_i}_{(P+1)} = {\overline{x}_i}_{(P+1)} + \Delta^0
\varepsilon_i(\vec{x}_{(P)}) + \Delta^1
\varepsilon_i(\vec{x}_{(P+1)}).
\end{equation}
\noindent

Why do we gain information? If we compare (\ref{pre0}) to (\ref{pre1}),
we can observe that the unknown term in (\ref{pre0}) is $\Delta^0
\varepsilon_i(\vec{x}_{(P+1)})$ which is (by hypothesis) much bigger
than the unknown term in (\ref{pre1}): $\Delta^1
\varepsilon_i(\vec{x}_{(P+1)})$. So, the term $\Delta^0
\varepsilon_i(\vec{x}_{(P)})$ is a correction to ${\overline{x}_i}_{(P+1)}$.
Analogously, we have $\Delta^2 \varepsilon_i(\vec{x}_{(P+1)}) =
\Delta^1 \varepsilon_i(\vec{x}_{(P+1)}) - \Delta^1
\varepsilon_i(\vec{x}_{(P)})$, implying that:
\begin{equation}
\label{del2}
\Delta^1 \varepsilon_i(\vec{x}_{(P+1)}) =  \Delta^1
\varepsilon_i(\vec{x}_{(P)}) + \Delta^2 \varepsilon_i(\vec{x}_{(P+1)}),
\end{equation}
\noindent
substituting this into (\ref{pre1}), if $(P)$ and $(P+1)$ are
sufficiently close such that ${\Delta^2
\varepsilon_i} \ll {\Delta^1 \varepsilon_i}$, we would have a second
order correction to  ${\overline{x}_i}_{(P+1)}$. Actually, when the
relation ${\Delta^{k+1} \varepsilon_i} \ll {\Delta^{k} \varepsilon_i}$
applies, we can further correct ${\overline{x}_i}_{(P+1)}$, i.e.,
\begin{widetext}
\begin{equation}
\label{prek}
{x_i}_{(P+1)} = {\overline{x}_i}_{(P+1)} + \Delta^0
\varepsilon_i(\vec{x}_{(P)}) + \Delta^1
\varepsilon_i(\vec{x}_{(P)}) + \cdots + \Delta^k
\varepsilon_i(\vec{x}_{(P)}) + \Delta^{k+1}
\varepsilon_i(\vec{x}_{(P+1)}).
\end{equation}
\end{widetext}
\noindent
Therefore, we can build a simple algorithm to improve the prediction
${\overline{x}_i}_{(P+1)}$, obtained with mapping ${\overline{M}}$: we
determine the integer $k$ for which the approximation starts to fail,
i.e., ${\Delta^{k+1} \varepsilon_i} \approx {\Delta^{k} \varepsilon_i}$,
then we neglect the term $\Delta^{k+1} \varepsilon_i(\vec{x}_{(P+1)})$
and end up with
\begin{eqnarray}
\label{preapprox}
{x_i}_{(P+1)} \cong {\overline{x}_i}_{(P+1)} + \Delta^0
\varepsilon_i(\vec{x}_{(P)}) + \nonumber \\
\Delta^1 \varepsilon_i(\vec{x}_{(P)}) + \cdots + \Delta^k
\varepsilon_i(\vec{x}_{(P)}).
\end{eqnarray}

The remaining question is: How to define ${\Delta^{k+1} \varepsilon_i}
\approx {\Delta^{k} \varepsilon_i}$? Let us elaborate the analysis just made
above. We are interested in using an approximation, a kind of Taylor series expansion,
when trying to forecast the Time Series, what one might expect from such
a situation? In a perfect world, the terms in the series would,
gradually, become smaller in an infinite fashion. Of course, as already
mentioned above, we are dealing with a real series, where each entry is
not infinitesimally apart the previous one and is, actually, finitely
separated. How ``finitely separated'' depends on the particular series
under study and, being more rigorous, on the particular section of the
series we are considering. This translates to the fact that, if one
considers the absolute values of the differences ${\Delta^{k}
\varepsilon_i}$, they will decrease with increasing values for $k$
until this value reaches the magnitude defined by the ``non-
infinitesimal'' character of the Time Series we have just emphasized,
where this character will then make the values for the differences
oscillate (for a while) around this magnitude (since this magnitude would
dominate over the initial tendency of the differences to decrease). With
the increasing values for $k$, this initial tendency of the differences
to decrease will cease as our approximation (Taylor like) stars to
diverge from the actual value for the series. We will then see the
absolute values for the following differences start to increase and
rapidly diverge. That clearly, if one thinks in plotting the (absolute)
values for the differences, defines a plateau where ${\Delta^{k+1} \varepsilon_i}
\approx {\Delta^{k} \varepsilon_i}$ and our above introduced method will
work at its best.

\subsection{The steps of the algorithm}

Consider that we have already reconstructed the phase space from the
Time Series under study and that we want to forecast the $P+1$ entry
($P$ is the last known value of the series). This entry corresponds to
a coordinate of a reconstructed vector on the phase space (as
usual). Using a global mapping $\overline{M}$, obtained via standard
k-fold validation procedures \cite{sarle}, we do the following:

\begin{enumerate}
\item Set $n=10$.
\item We calculate the absolute value for the functions
$\Delta^k\varepsilon_i$  (see eq.(\ref{varepReal})) up to $k=n$ for the point $P$.
\item We check to see if we have already found the plateau, i.e., we
look for the value of $k$ for which $|\Delta^k\varepsilon_i| <
|\Delta^{k+1}\varepsilon_i|$. Please note that this checking can be very
easily automatized.
\item
If the checking returns false we set
n=n+10 and return to step 2. Otherwise we would have found the corrected
value for ${x_i}_{(P+1)}$ as:
\begin{eqnarray}
\label{preapprox}
{x_i}_{(P+1)} \cong {\overline{x}_i}_{(P+1)} + \Delta^0
\varepsilon_i(\vec{x}_{(P)}) + \nonumber \\
\Delta^1
\varepsilon_i(\vec{x}_{(P)}) + \cdots + \Delta^k
\varepsilon_i(\vec{x}_{(P)}).
\end{eqnarray}
\end{enumerate}
\section{Applications}

In this section, we are going to present two applications of the above
introduced improved forecast method. We will start by introducing the
Time Series in question, present the reconstruction parameters and the
associated Global Mapping. We then will proceed to the algorithm,
following the steps just introduced and compare the average performance
for the ``usual'' and the improved approaches.

\subsection{Application 1: Lorenz}
\label{app1}

\subsubsection{The Time Series}

This is an ``academic'' application in the sense that it is, certainly,
originated from a dynamic system and we actually even know which one.
But it is important in order for us to see the ideas of the improved
method working on an arena that suits it very nicely.

\begin{figure}[p]
\begin{center}
\includegraphics[width=7cm,angle=-90]{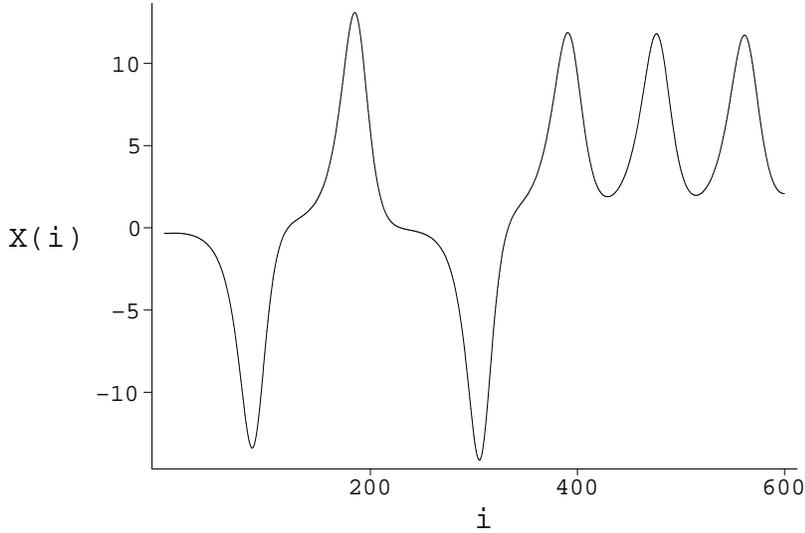}
\end{center}
\caption{Lorenz Time series. The horizontal axis marks the
position of the entry (i) and the vertical on the value for the
entry (X(i))} \label{LORENZ}
\end{figure}

The Time Series was generated taking the consecutive values for
the $x_1$ coordinate of the Lorenz system (see eq.
(\ref{sistLorenz})), starting from the initial condition
$x_{1_0}=-0.3336666667,x_{2_0}=-0.3336666667,x_{3_0}=21.9996666667$,
using an eighth order Runge-Kutta numerical
integration\cite{ndyn}. The Series presents 600 entries (please
see figure (\ref{LORENZ}) for a plotting of this Time Series).

Now, in order to apply the Global Analysis
(\cite{AbarbanelN,Abarbanel}) to this Time Series, we have to
reconstruct the phase space. To do that, we need to determine the
relevant parameters, namely the time-lag and the embedding
dimension (please see \cite{embd}). For this present case, the
reconstruction parameters are time-lag = 6 and embedding dimension
= 3. So, in the remaining of this subsection, we will call these
three dimensions of the reconstructed phase space for the Lorenz
system $(x,y,z)$. In real life, we use the whole Time Series we
know/measure to produce the Global Mapping and use it to predict
future (unknown) entries. Here, in order to evaluate the accuracy
of the predictions we obtain using a regular Global Fitting and
our Improved one, we are going to use an initial portion of the
Series to generate the Mapping and the other (remaining) portion
of the Series as our testing ground, i.e., we will apply our
mappings to entries in that region and compare it to the actual
values to see how the mappings fared. In the present case, the
first 140 entries constitute our portion of the Series used to
build the Mapping up. Basically, we use all the vector
reconstructed from these entries and produce a quadratic fitting
minimizing the distances from this fitting (when applied to each
vector) to the actual values via, for instance, a least mean
square procedure. Actually, we also have used an improvement (a
very standard one) called a k-validation. In layman's language,
basically what this k- validation does is to average up several
mappings. Doing all this, the global mapping we have derived (and
to be used on this application henceforth) is:
\begin{eqnarray}
\label{Mlorenz}
\overline{M} = 1.317833301\,x- 0.005266089766\,{x}^{2} \nonumber \\
+ 0.07580676400\,xy  -  0.1245478927\,xz -  \nonumber \\
0.01839588238\,{y}^{2} + 0.06578287850\,yz -  \nonumber \\
0.01025562766\,{z}^{2} - 0.4700554502\,y +  \nonumber \\
0.1415056465\,z.  \nonumber \\
\bigskip
\end{eqnarray}

\subsubsection{The inner works of the improved forecast algorithm}

Let us now, using two generic points from the Series, exemplify the
workings of our improved method.

Consider the entries $P = 316$ and $P = 533$, with respective values of
$-1.370578116$ and $6.860383245$. The values for the entries $P = 317$ and
$P = 534$, the ``next'' entry for each case considered here, are $-1.041455029$ and $7.225654731$. Let us see how the ``usual'' Global
Fitting fares in these entries. Using the mapping presented on
(\ref{Mlorenz}), we get the following forecasting for the entries $P = 317$ and
$P = 534$: $-0.782644049$ and $7.062374264$. These present a
``percentage error'' (given by $|(value - forecast)/value|$) of $24.85090309$ and $2.259732482$ respectively.
How about the improved method?

In order to apply our method we have to find the plateau by finding
the value for $k$ to which $|\Delta^k\varepsilon_i| <
|\Delta^{k+1}\varepsilon_i|$. Let us do that for the couple of points
chosen above:

\begin{itemize}
\item P=316

As ``prescribed'' above, what we have to do is, by looking at table
(\ref{lorenz_plateau_316}), second column, determine at which value of $k$
$\Delta^k\varepsilon(\vec{x}_{(P)})$ stops decreasing for the first
time (and begin the oscillations we have mentioned in section \ref{dalgo}).
From table (\ref{lorenz_plateau_316}), we see that happens for $k=5$. Using this
into equation (\ref{preapprox}), we find (see table (\ref{lorenz_plateau_316})) that the
``percentage error'' for our method is 0.0004798095.

\item P=533
Again, what we have to do is, by looking at table
(\ref{lorenz_plateau_533}), second column, determine to which value of $k$
$\Delta^k\varepsilon(\vec{x}_{(P)})$ stops decreasing for the first
time (and begin the oscillations we have mentioned in section \ref{dalgo}).
From table (\ref{lorenz_plateau_533}), we see that happens for $k=3$. Using this
into equation (\ref{preapprox}), we find (see table (\ref{lorenz_plateau_533})) that the
``percentage error'' for our method is 0.0001986533.
\end{itemize}

\begin{figure}[p]
\begin{center}
\includegraphics[width=9cm,height=8cm,angle=-90]{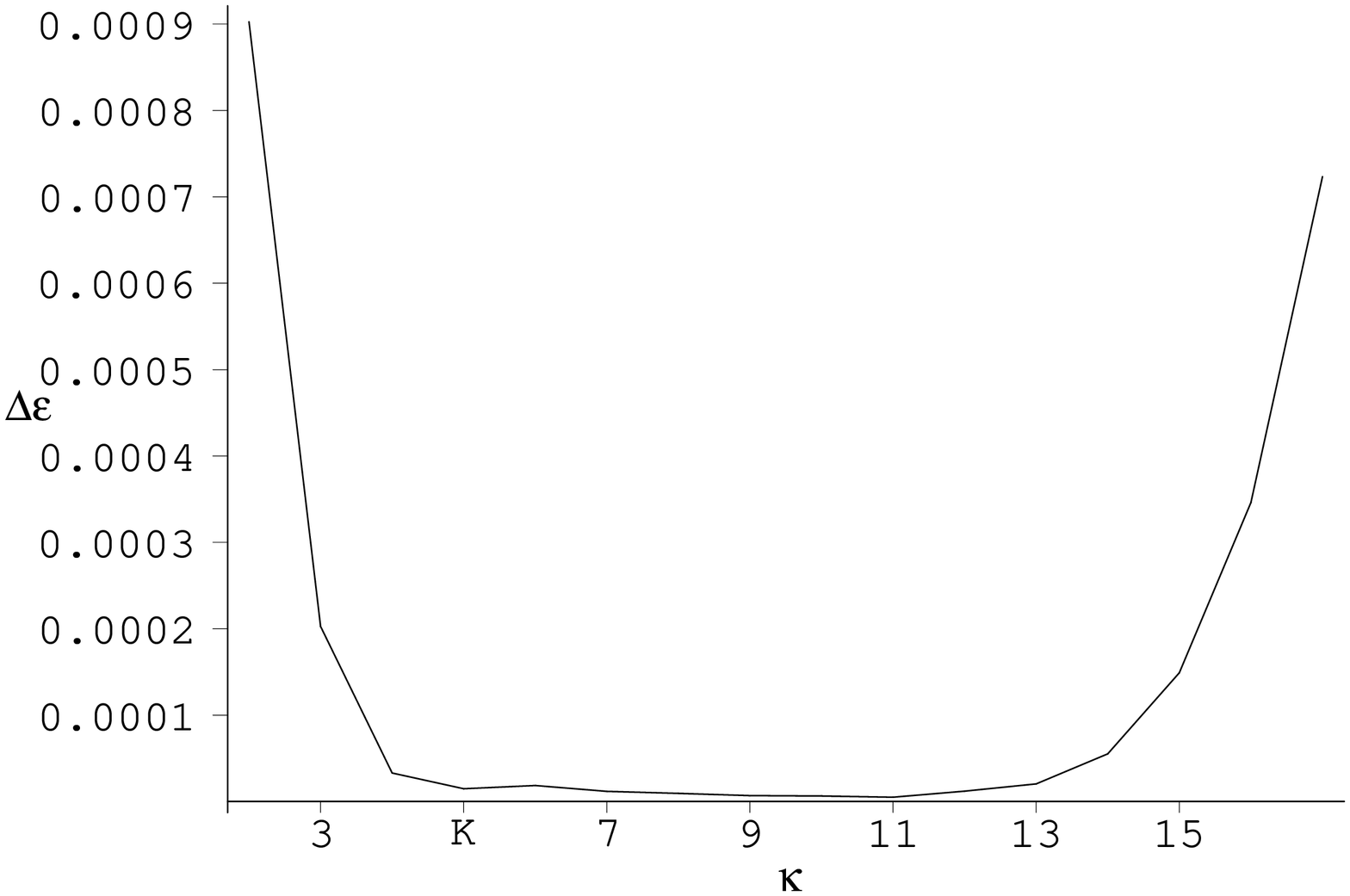}

\caption{The plot shows the values of the $|\Delta^k\varepsilon|$ against the number $k$, for the entry 316, for the
Lorenz System Time Series. In the x-axis, marked with the letter $K$, is the value of $k$ that our procedure
defines as the beginning of the plateau.}
\label{316}
\end{center}
\end{figure}

\begin{figure}[p]
\begin{center}
\includegraphics[width=7cm,height=8cm,angle=-90]{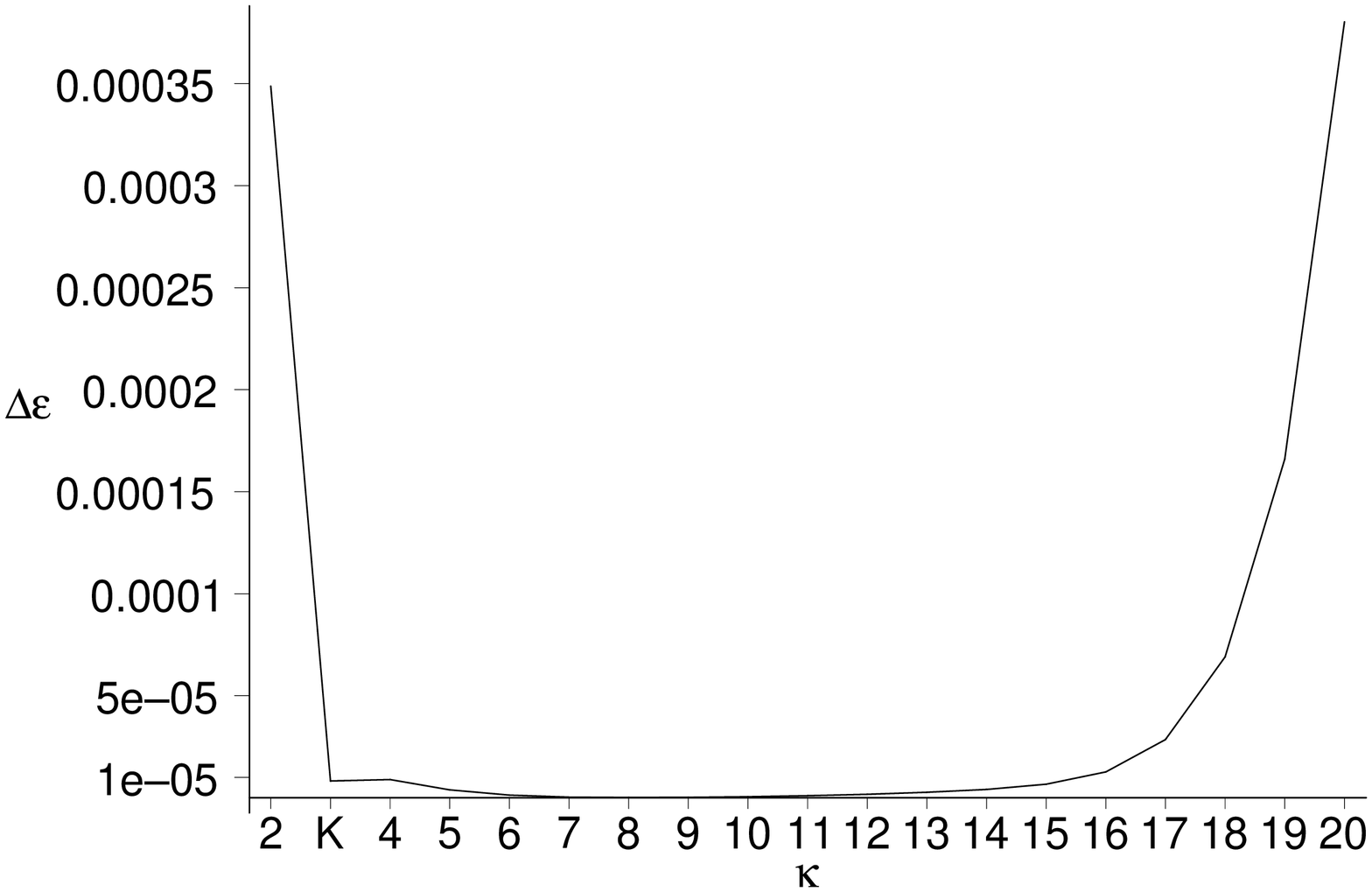}
\caption{The plot shows the values of the $|\Delta^k\varepsilon|$ against the value of $k$, for the entry 533, for the
Lorenz System Time Series. In the x-axis, marked with the letter $K$, is the value of $k$ that our procedure
defines as the beginning of the plateau.}
\label{533}
\end{center}
\end{figure}

\bigskip
\bigskip
As we have mentioned in section (\ref{dalgo}), we expect the absolute values
of $\Delta^k\varepsilon(\vec{x}_{(P)})$ to oscillate when
$|\Delta^k\varepsilon_i| \approx |\Delta^{k+1}\varepsilon_i|$. That fact
is illustrated, for the entries $P=316$ and $P=533$ respectively, on
figures (\ref{316}) and (\ref{533}).

\subsubsection{Performance Comparison}

The reader may ask: why these two entries above? Fair enough, they are
not special at all. So, in order to confirm the fact that our new
approach may be an advantage, let us make a general survey of the
entries on the Time Series. We take 21 entries, equally
distributed, on the last part (not used when producing the Global
Mapping) of the Time Series. The results are
presented on table (\ref{Comparison_Lorenz}).

The idea behind of presenting the results for points equally
spaced on the entire Time Series (meaning the entire testing
ground defined above) was to provide the information on all the
Time Series, i.e., it is very important (for many Series) the
section in which the analysis is carried out. So, we have decided
to present the results for many points, evenly distributed along
every section of the Time Series. But, for completeness, we will
present the average percentage error (for the improved Global
fitting) for the whole testing ground for the Time Series and for
the 21 entries used on table (\ref{Comparison_Lorenz}). The
percentage error for the whole series is $.1601961683e-1$ and for the 21 entries
on the table is $.3385849199e-1$. Both are compatible, showing that the chosen
21 are representative of the totality of the possibilities. The
percentage error for the ``regular'' global fitting is (for the whole series) $10.58939930$.

\begin{figure}[p]
\begin{center}
\includegraphics[width=7cm,height=8cm,angle=-90]{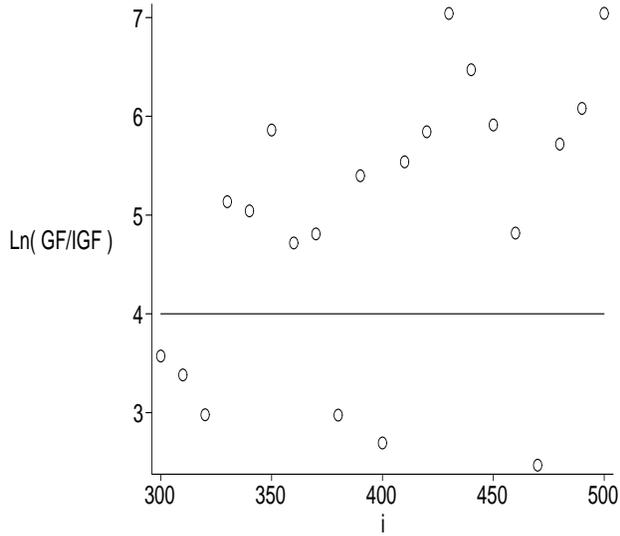}
\caption{The plot is for ln($\Delta_{GF} /
\Delta_{IGF})$ against the position in the Time Series ($i$). The line marks the threshold where, above it,
$\Delta_{IGF}$ starts to be smaller than
$e^{-4}$ times $\Delta_{GF}$.}
\label{lorenz_ln}
\end{center}
\end{figure}

\bigskip
\bigskip
As can be seen, our method is a great improvement of accuracy when compared
with the ``plain'' Global Fitting. To help in this analysis we present
figure (\ref{lorenz_ln}) where we plot $\ln(\Delta_{GF} / \Delta_{IGF})$, where $\Delta_{GF}$
and $\Delta_{IGF}$ are, respectively the percentage errors in the Global Fitting and
the Improved Global Fitting. As can be seen from the figure, most of the
IGF errors are smaller than $e^{-4}$ times the GF errors.

\subsection{Application 2: Heart beat}
\label{app2}

\subsubsection{The Time Series}

\begin{figure}[p]
\begin{center}
\includegraphics[width=7cm,height=8cm,angle=-90]{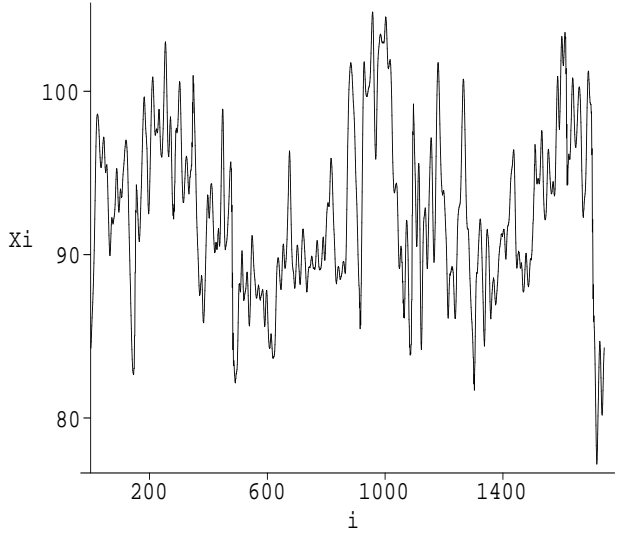}
\caption{Heartbeat data. The horizontal axis marks the position of
the entry (i) and the vertical on the value for the entry (X(i))}
\label{HeartBeaT}
\end{center}
\end{figure}

Let us now deal with a more ``real'' example, where we deal with data
extracted from Nature, we do not know the system behind the phenomenon,
etc. The following Time Series was obtained \footnote{http://ecg.mit.edu/time-series/} from measurements of the
heart beat rate in a person performing many different activities. The
Series presents 1744 entries (please see figure (\ref{HeartBeaT}) for a plotting of this Time Series).

In order to produce the Global Mapping for this case, we have
proceeded in the same fashion as we did in the Lorenz System Time
Series application. So, we will not repeat the whole explanation
of the procedures involved here. Please refer to section IV-A
above. For this application, the reconstruction parameters are
time-lag = 10 and embedding dimension = 3. So, in the remaining of
this subsection, we will call these three dimensions of the
reconstructed phase space for the Heart beat data $(x,y,z)$. The
global mapping we have derived (and to be used on this application
henceforth) is:
\begin{eqnarray}
\label{Mheart}
\overline{M} = - 1.172534275\,x- 0.2617292220\,{z}^{2} +  \nonumber \\
0.4661889468\,yz -  0.3426537822\,{y}^{2} -  \nonumber \\
0.1107944588\,xz+ 0.3574412776\,xy -   \nonumber \\
 0.1136908621\,{x}^{2}+ 15.65933419\,z -  \nonumber \\
12.98866231\,y. \nonumber \\
\bigskip
\end{eqnarray}

\subsubsection{The inner works of the improved forecast algorithm}

Let us now, using two generic points from the Series, exemplify the
workings of our improved method.

As in the previous application, consider the entries $P = 737$ and $P =
1016$, with respective values of $89.18875624$ e $94.25098125$. The
values for the entries $P = 738$ and $P = 1017$, the ``next'' entry for
each case considered here, are 89.16743126 and 94.28981563. Let us see
how the ``usual'' Global Fitting fares in these entries. Using the
mapping presented on (\ref{Mheart}), we get the following forecasting
for the entries $P = 738$ and $P = 1017$: 90.996977 and 82.611004. These present a ``percentage error'' (defined above) of
$2.051809404$ and $12.38607961$ respectively. How about the improved
method?

In order to apply our method we have to find the plateau by finding
the value for $k$ to which $|\Delta^k\varepsilon_i| <
|\Delta^{k+1}\varepsilon_i|$. Let us do that for the couple of points
chosen above:

\begin{itemize}

\item{P=737}

As ``prescribed'' above, what we have to do is, by looking at table
(\ref{delta_737}), second column, determine to which value of $k$
$\Delta^k\varepsilon(\vec{x}_{(P)})$ stops decreasing for the first
time (and begin the oscillations we have mentioned in section \ref{dalgo}).
From table (\ref{delta_737}), we see that happens for $k=1$. Using this
into equation (\ref{preapprox}), we find (see table (\ref{delta_737})) that the
``percentage error'' for our method is 0.3970473916.

\bigskip
\bigskip
\item P=1016

Again, what we have to do is, by looking at table
(\ref{delta_1016}), second column, determine to which value of $k$
$\Delta^k\varepsilon(\vec{x}_{(P)})$ stops decreasing for the first
time (and begin the oscillations we have mentioned in section \ref{dalgo}).
From table (\ref{delta_1016}), we see that happens for $k=3$. Using this
into equation (\ref{preapprox}), we find (see table (\ref{delta_1016})) that the
``percentage error'' for our method is 1.524959626.

\end{itemize}

\begin{figure}[p]
\begin{center}
\includegraphics[width=7cm,height=8cm,angle=-90]{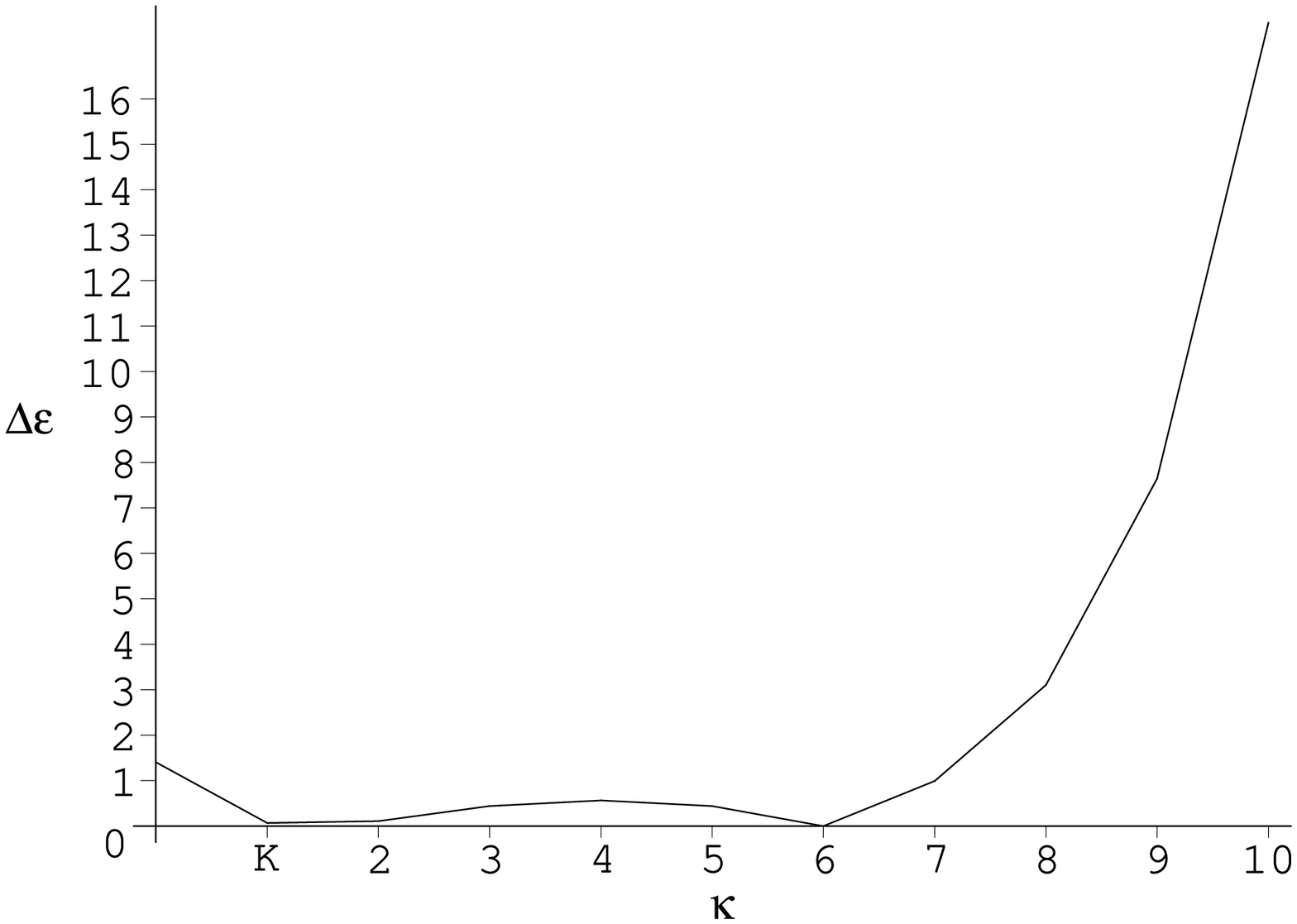}
\caption{The plot shows the values of the $|\Delta^k\varepsilon|$ against the value of $k$, for the entry 737, for the
Heartbeat Time Series. In the x-axis, marked with the letter $K$, is the value of $k$ that our procedure
defines as the beginning of the plateau.}
\label{737}
\end{center}
\end{figure}

\begin{figure}[p]
\begin{center}
\includegraphics[width=7cm,height=8cm,angle=-90]{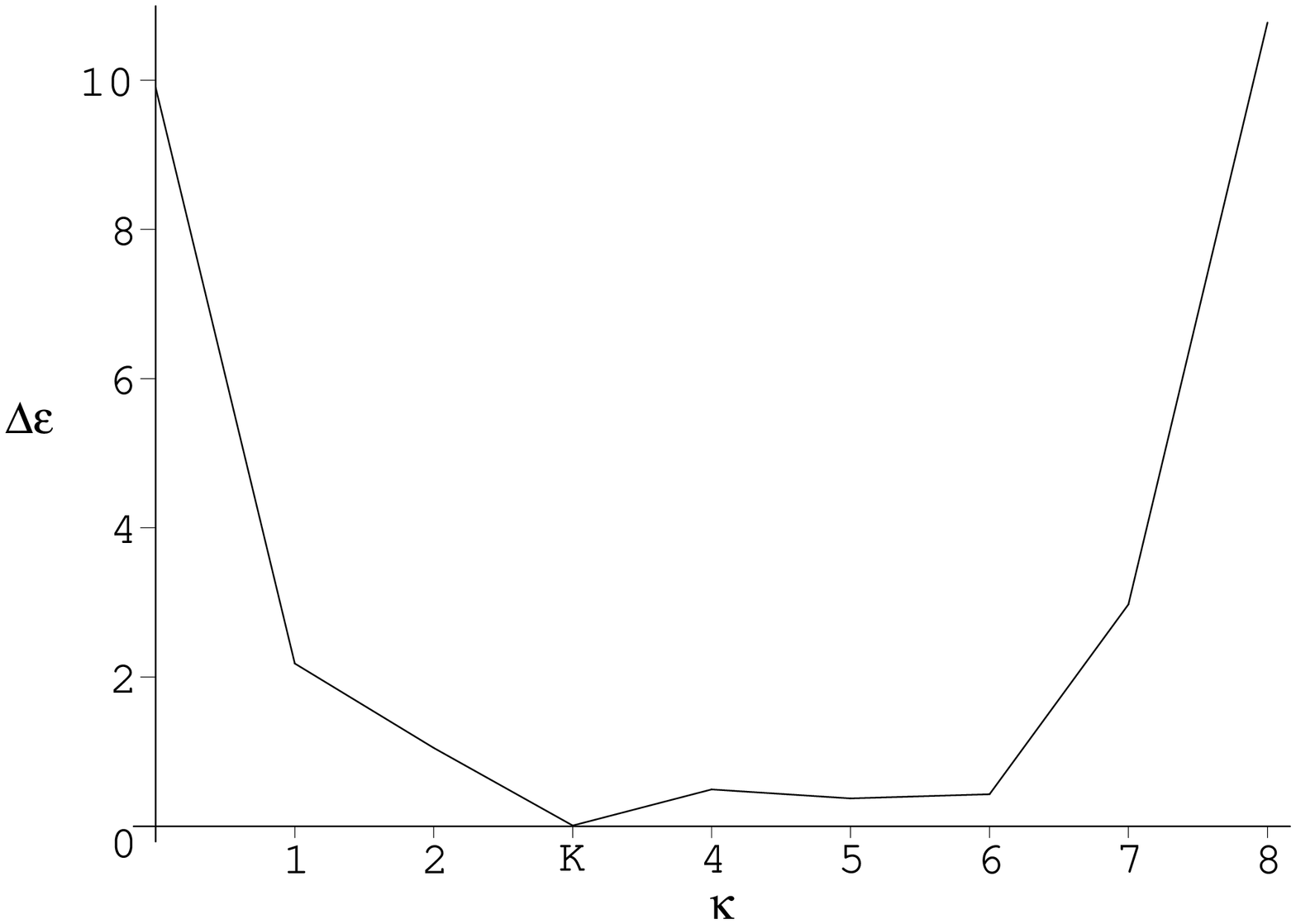}
\caption{The plot shows the values of the $|\Delta^k\varepsilon|$ against the vale for $k$, for the entry 1016, for the
Heartbeat Time Series. In the x-axis, marked with the letter $K$, is the value of $k$ that our procedure
defines as the beginning of the plateau.}
\label{1016}
\end{center}
\end{figure}

As in the previous application, we expect the absolute values
of $\Delta^k\varepsilon(\vec{x}_{(P)})$ to oscillate when
$|\Delta^k\varepsilon_i| \approx |\Delta^{k+1}\varepsilon_i|$. That fact
is illustrated, for the entries $P=737$ and $P=1016$ respectively, on
figures (\ref{737}) and (\ref{1016}).

\subsubsection{Performance Comparison}

Let us make the general survey of the
entries on this Time Series. We take 26 entries, equally
distributed, on the last part (not used when producing the Global
Mapping) of the Time Series. The results are
presented on table (\ref{Comparison_Heart}).

The idea behind of presenting the results for points equally
spaced on the entire Time Series (meaning the entire testing
ground defined above) is the same one explained on the section
regarding the Lorenz System. The percentage error for the whole
series is $1.877994467$ and for the 26 entries on the table is
$1.429515613$. Both are compatible, showing that the chosen 26 are
representative of the totality of the possibilities. The
percentage error for the ``regular'' global fitting (for the whole
series) is $5.971546764$.

\begin{figure}[p]
\begin{center}
\includegraphics[width=7cm,height=8cm,angle=-90]{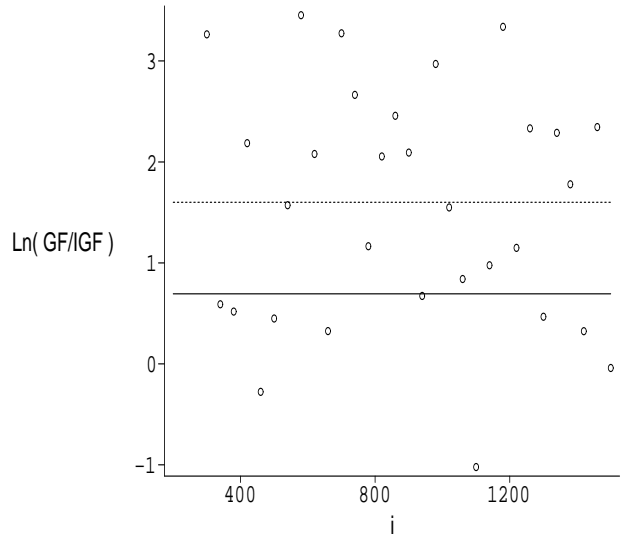}
\caption{The plot is for ln($\Delta_{GF} / \Delta_{IGF})$ against the position in the Time Series ($i$). The solid line marks the threshold where, above it,  $\Delta_{IGF}$ starts to
be the half of $\Delta_{GF}$ and the dotted line the threshold where, above it, $\Delta_{IGF}$ starts to
be be the fifth of $\Delta_{GF}$.}
\label{heart_ln}
\end{center}
\end{figure}

\bigskip
\bigskip
As can be seen, in the majority of cases, our method is, for this more
``realistic'' case, also a great improvement of accuracy when compared
with the ``plain'' Global Fitting. To help in this analysis we present
figure (\ref{heart_ln}) where we plot $\ln(\Delta_{GF} / \Delta_{IGF})$, where $\Delta_{GF}$
and $\Delta_{IGF}$ are, respectively the percentage errors in the Global Fitting and
the Improved Global Fitting. As can be seen from the figure, most of the
IGF errors are more than five times smaller than the GF errors.
\section{Conclusion}
\label{conclu}

There is a huge demand for improving methods that do not cost too
high a computational price to achieve desired levels of accuracy
in Time Series Analysis.

Here, we have presented one such method. The basic rational behind
it is that we can make use, as explained in section \ref{algo}, of
the underlying (assumed) low-dimensionality dynamics to correct
our forecast. It is important to mention that, in order to apply
the method, one does not have to quantify the hyperbolicity (or
the low-dimensionality, for that matter) of the Time Series. The
steps of the procedure will take (automatically) care of stopping
when this hyperbolicity ``spoils'' the correcting power of the
method. So, the algorithm is secure. It is also useful to remember
that our efforts here are aimed to avoid the computational cost of
the fitting/minimizing procedures. So, our method is not
equivalent to fittings, with the same computational cost, in any
shape or form.

We have presented two applications of our method: The first one is a (we
are going to call it) pure low dimensional known system, from where we
generated a Time Series. The reason for this application is to use the
method on a controlled arena, i.e., we can see the method working at its
best. What do we mean by its best? Could not have we gotten better
results than the ones presented in section \ref{app1}? Of course we could
have, for instance, if we have made the Time Series more ``dense'', i.e.
if we have used smaller values for $\Delta t$, of course, the results
would be better. Indeed, we can do the same indefinitely up to infinite
precision. What we mean by ``its best'' is the fact that there is not,
for sure, any high dimensional behavior. We have then demonstrated that
the ideas behind our method work quite nicely.

The second application corresponds to a Time Series obtained from
measurements, i.e., we do not have any prior knowledge about the
(possible) dynamic system underlying it. We have found that, after
the usual techniques have been used to produce the Global mapping,
we could improve the forecast capabilities of the fitting quite a
bit (see section \ref{app2}), thus demonstrating the practicality
of our approach on a uncontrolled situation.

Our method has, of course, its limitations. Perhaps the most
obvious one is the fact that it won't help much in the case where
the Time Series is ``sparse'', i.e., as we have mentioned just
above, as $\Delta t$ becomes large, the method won't work. The
limitation so far is that we do not have a criteria, as yet, to,
just by quickly inspecting the Time Series, determine if our
method applies well or not. One has to have a go and, in a testing
arena, verify if the method is improving things.

That leads to future work: produce a fast algorithm to test the time
series for applicability (or not) of the method. One other possible line
of research to be pursued is to improve our algorithm in the sense of
using more information contained on the plateau than we are using now.
So far, we are taking the first piece of data on the plateau but, as
we have explained in section \ref{dalgo}, the values for the corrections will
oscillate from that point on. It is reasonable to look for an algorithm
to extract information from this oscillation.


\begin{table}[p]
\begin{center}
\begin{tabular}{|l|l|l|l|} \hline
$k$ & $|\Delta^k\varepsilon(\vec{x}_{(P)})|$ & IGF & error \\ \hline
1& 0.015127123&- 1.042168004& 0.06845950907\\ \hline
2& 0.000902659&- 1.041265345& 0.01821336445\\ \hline
3& 0.000202785&- 1.041468130& 0.001257951581\\ \hline
4& 0.000032905&- 1.041435225& 0.001901570346\\ \hline
5& 0.000014807&- 1.041450032& 0.0004798094839\\ \hline
6& 0.000018502&- 1.041468534& 0.001296743462\\ \hline
7& 0.000011662&- 1.041456872& 0.0001769639541\\ \hline
8& 0.000009405&- 1.041447467& 0.0007260995232\\ \hline
9& 0.000006933&- 1.041454400& 0.00006039627084\\ \hline
10& 0.000006450&- 1.041460850& 0.0005589295589\\ \hline
11& 0.000005072&- 1.041455778& 0.00007191861186\\ \hline
12& 0.000011998&- 1.041443780& 0.001080123451\\ \hline
13& 0.000020272&- 1.041423508& 0.003026630927\\ \hline
14& 0.000055212&- 1.041368296& 0.008328060030\\ \hline
15& 0.000149033&- 1.041219263& 0.02263813544\\ \hline
16& 0.000346170&- 1.040873093& 0.05587720869\\ \hline
17& 0.000723317&- 1.040149776& 0.1253297515\\ \hline
18& 0.001398781&- 1.038750995& 0.2596400156\\ \hline
19& 0.002499815&- 1.036251180& 0.4996710232\\ \hline
20& 0.004042167&- 1.032209013& 0.8877979118\\ \hline
\end{tabular}
\end{center}
\caption{In this table, we plot the $|\Delta^k\varepsilon|$, for the entry 316, for the Lorenz System Time Series. IGF is our improved global fitting result corresponding to the particular value of $k$.}
\label{lorenz_plateau_316}
\end{table}

\begin{table}[p]
\begin{center}
\begin{tabular}{|l|l|l|l|} \hline
$k$ & $\Delta^k\varepsilon(\vec{x}_{(P)})$ & IGF & error \\ \hline
1& 0.005833355& 7.225283437& 0.005138551644\\ \hline
2& 0.000348756& 7.225632193& 0.0003119163708\\ \hline
3& 0.000008184& 7.225640377& 0.0001986532783\\ \hline
4& 0.000008912& 7.225649289& 0.00007531497425\\ \hline
5& 0.000003874& 7.225653163& 0.00002170045565\\ \hline
6& 0.000001257& 7.225654420& 0.000004304108231\\ \hline
7& 0.000000331& 7.225654751& 0.0000002767915261\\ \hline
8& 0.000000150& 7.225654901& 0.000002352727972\\ \hline
9& 0.000000251& 7.225655152& 0.000005826461624\\ \hline
10& 0.000000532& 7.225655684& 0.00001318911622\\ \hline
11& 0.000001007& 7.225656691& 0.00002712556956\\ \hline
12& 0.000001712& 7.225658403& 0.00005081892419\\ \hline
13& 0.000002692& 7.225661095& 0.00008807506360\\ \hline
14& 0.000004115& 7.225665210& 0.0001450249201\\ \hline
15& 0.000006683& 7.225671893& 0.0002375148085\\ \hline
16& 0.000012706& 7.225684599& 0.0004133604651\\ \hline
17& 0.000028487& 7.225713086& 0.0008076084753\\ \hline
18& 0.000068998& 7.225782084& 0.001762511561\\ \hline
19& 0.000166009& 7.225948093& 0.004060005784\\ \hline
20& 0.000380304& 7.226328397& 0.009323252011\\ \hline
\end{tabular}
\end{center}
\caption{In this table, we plot the $|\Delta^k\varepsilon|$, for the entry 533, for the Lorenz System Time Series. IGF is our improved global fitting result corresponding to the particular value of $k$.}
\label{lorenz_plateau_533}
\end{table}

\begin{table}[p]
\begin{center}
\begin{tabular}{|l|l|l|} \hline
$N$ & GF error & IGF error \\ \hline
300& 2.949988321& 0.0007879476836\\ \hline
310& 9.384955078& 0.003895726200\\ \hline
320& 602.9300615& 0.6329500170\\ \hline
330& 4.588145520& 0.00003360765471\\ \hline
340& 0.6804060144& 0.000006172635711\\ \hline
350& 1.799833141& 0.000002478938512\\ \hline
360& 1.908426262& 0.00003660198120\\ \hline
370& 1.992750955& 0.00003095127924\\ \hline
380& 5.351353323& 0.005676508421\\ \hline
390& 15.14897504& 0.00006055593702\\ \hline
400& 17.40670926& 0.03531034693\\ \hline
410& 11.79516640& 0.00003410572689\\ \hline
420& 6.417051601& 0.000009213921336\\ \hline
430& 3.561881518& 0.0000003239205212\\ \hline
440& 2.472699706& 0.0000008353740914\\ \hline
450& 2.070219097& 0.000002536966462\\ \hline
460& 2.807205469& 0.00004274128369\\ \hline
470& 9.466251097& 0.03219442293\\ \hline
480& 17.78540068& 0.00003404766212\\ \hline
490& 15.54182977& 0.00001293826149\\ \hline
500& 9.424552928& 0.0000008546217942\\ \hline
\end{tabular}
\end{center}
\caption{Comparison between the Global Fitting and the Improved Global Fitting for the Lorenz Time Series}
\label{Comparison_Lorenz}
\end{table}

\begin{table}[p]
\begin{center}
\begin{tabular}{|l|l|l|l|} \hline
$k$ & $\Delta^k\varepsilon(\vec{x}_{(P)})$ & IGF & error \\ \hline
0& 1.40709476& 89.58988224& 0.4737727375\\ \hline
1& 0.06841402& 89.52146822& 0.3970473916\\ \hline
2& 0.11034922& 89.63181744& 0.5208024650\\ \hline
3& 0.44228546& 90.07410290& 1.016819288\\ \hline
4& 0.56575633& 90.63985923& 1.651306928\\ \hline
5& 0.43932335& 91.07918258& 2.144001787\\ \hline
6& 0.00174856& 91.07743402& 2.142040802\\ \hline
7& 0.99434282& 90.08309120& 1.026899538\\ \hline
8& 3.10231730& 86.98077390& 2.452304983\\ \hline
9& 7.64654504& 79.33422886& 11.02779598\\ \hline
10& 17.67633668& 61.65789218& 30.85155498\\ \hline
\end{tabular}
\end{center}
\caption{In this table, we plot the $|\Delta^k\varepsilon|$, for the entry 737, for the Time Series with the Heart Beat data}
\label{delta_737}
\end{table}

\begin{table}[p]
\begin{center}
\begin{tabular}{|l|l|l|l|} \hline
$k$ & $\Delta^k\varepsilon(\vec{x}_{(P)})$ & IGF & error \\ \hline
0& 9.89683125& 92.50783525& 1.889896982\\ \hline
1& 2.18086125& 94.68869650& 0.4230370664\\ \hline
2& 1.05010575& 95.73880225& 1.536737144\\ \hline
3& 0.01110500& 95.72769725& 1.524959626\\ \hline
4& 0.49314562& 95.23455163& 1.001949143\\ \hline
5& 0.37340972& 94.86114191& 0.6059257579\\ \hline
6& 0.43065248& 95.29179439& 1.062658521\\ \hline
7& 2.97396166& 98.26575605& 4.216723082\\ \hline
8& 10.77153154& 109.0372876& 15.64057780\\ \hline
9& 32.44747527& 141.4847629& 50.05306984\\ \hline
10& 86.66665506& 228.1514179& 141.9682512\\ \hline
\end{tabular}
\end{center}
\caption{In this table, we plot the $|\Delta^k\varepsilon|$, for the entry 1016, for the Time Series with the Heart Beat data}
\label{delta_1016}
\end{table}

\begin{table}[p]
\begin{center}
\begin{tabular}{|l|l|l|} \hline
$N$ & GF error & IGF error \\ \hline
500& 3.397125034& 2.168865202\\ \hline
540& 9.089213645& 1.889289807\\ \hline
580& 5.874155830& 0.1858455792\\ \hline
660& 1.844997944& 1.332964363\\ \hline
700& 0.3583492837& 0.01355789862\\ \hline
740& 2.400959523& 0.1672692403\\ \hline
780& 0.4937771060& 0.1540773590\\ \hline
820& 1.249041343& 0.1600760146\\ \hline
860& 10.26039663& 0.8787867282\\ \hline
900& 6.962087707& 0.8589041386\\ \hline
980& 4.565823761& 0.2342961164\\ \hline
1020& 14.67475919& 3.116867623\\ \hline
1060& 1.249489572& 0.5393005133\\ \hline
1100& 1.075608307& 2.987879519\\ \hline
1140& 0.2169661915& 0.08177074130\\ \hline
1180& 13.66135448& 0.4848516873\\ \hline
1220& 2.732512685& 0.8664583752\\ \hline
1260& 6.861097043& 0.6660946812\\ \hline
1340& 9.400776847& 0.9535935739\\ \hline
1380& 1.392042607& 0.2352261816\\ \hline
1420& 0.8176069275& 0.5917808721\\ \hline
1460& 2.813583072& 0.2695624030\\ \hline
1500& 6.384669758& 6.651398095\\ \hline
\end{tabular}
\end{center}
\caption{Comparison between the Global Fitting and the Improved Global Fitting for the Time Series with Heart Beat data}
\label{Comparison_Heart}
\end{table}

\end{document}